\documentclass[12pt]{article}
\usepackage{latexsym}
\title{Proof that Casimir force does not originate from vacuum energy}
\author{Hrvoje Nikoli\'c \\
Theoretical Physics Division, Rudjer Bo\v{s}kovi\'{c} Institute, \\
P.O.B. 180, HR-10002 Zagreb, Croatia \\
{\normalsize e-mail: hnikolic@irb.hr} \\
\makebox[1in]{} \\
}
\date{\today}
\begin{document}
\maketitle
\begin{abstract}
We present a simple general proof that Casimir force cannot originate from 
the vacuum energy of electromagnetic (EM) field. The full QED Hamiltonian consists of 3 terms: 
the pure electromagnetic term $H_{\rm em}$, the pure matter term $H_{\rm matt}$ 
and the interaction term $H_{\rm int}$. 
The $H_{\rm em}$-term commutes with all matter fields because it
does not have any {\em explicit} dependence on matter fields.
As a consequence, $H_{\rm em}$ cannot generate 
any forces on matter. Since it is precisely this term that generates the vacuum energy of EM field,
it follows that the vacuum energy does not generate the forces. 
The misleading statements in the literature that vacuum energy generates Casimir force 
can be boiled down to the fact that $H_{\rm em}$ attains an {\em implicit} dependence 
on matter fields by the use of the equations of motion 
and to the illegitimate treatment of the implicit dependence as if it was explicit.  
The true origin of the Casimir force is van der Waals force generated by $H_{\rm int}$.   
\end{abstract}

\vspace*{0.5cm}
PACS Numbers: : 11.10.Ef, 03.70.+k, 42.50.Lc \newline


\section{Introduction}

The Casimir force \cite{casimir} is widely viewed as a force that originates 
from the vacuum energy, which is a view especially popular in the community of high-energy physicists 
\cite{itz-zub,BD,zin-just,mukhanov,zee}.
Another view, more popular in the condensed-matter community, is that Casimir force
has the same physical origin as van der Waals force 
\cite{casimir-polder,lifshitz,lifshitz2,LL9,mostepanenko-rmp,mostepanenko-advances,commins},
which does not depend on energy of the vacuum.
From a practical perspective, the two points of view appear as two complementary approaches,
each with its advantages and disadvantages.

From a fundamental perspective, however, one may be interested to know which of the two approaches
is more fundamental. After all, the conceptual picture of the world in which the vacuum energy 
has a direct physical role is very different from the picture in which it does not. 
From such a fundamental perspective, Jaffe argued \cite{jaffe} 
that the physically correct approach is the one based on van der Waals force, while the 
approach based on vacuum energy is merely a heuristic shortcut valid only as an approximation
in the limit of infinite fine structure constant. 
Similar doubts about the vacuum-energy approach to Casimir force has been expressed by Padmanabhan 
\cite{padmanabhan}. Nevertheless, it seems that a general
consensus is absent \cite{lamoreaux-pt,brevik-milton}. The question of relevance of the vacuum 
energy for Casimir force is still a source of controversy.

In this paper we present a theoretical way to resolve the controversy. 
In short, similarly to Jaffe \cite{jaffe},
we find that the approach based on vacuum energy is unjustified from a fundamental 
theoretical perspective, leaving only the non-vacuum van der Waals-like approaches as physically 
viable. However, to arrive at that conclusion, we use an approach very different 
from the approach used by Jaffe. Our approach is rather mathematical in spirit,
because our central idea is to carefully distinguish 
explicit dependence from implicit dependence in canonical equations of motion for 
classical and quantum physics. In this way our approach is more abstract 
and more general than the approach by Jaffe, but still sufficiently simple 
to be accessible to a wide readership of theoretical physicists.   

\section{Heuristic idea}

Let us start with a brief overview of the standard calculation 
\cite{itz-zub} of Casimir force 
from vacuum energy. The energy of electromagnetic (EM) field is
\begin{equation}\label{H1}
H_{\rm em}=\int d^3x\, \frac{{\bf E}^2+{\bf B}^2}{2} .
\end{equation}
In general, the fields ${\bf E}$ and ${\bf B}$ have Fourier transforms with contributions from 
all possible wave vectors ${\bf k}$. However, in the absence of electric currents in a conductor,
the use of Maxwell equations implies that the EM field 
must vanish at conducting plates. Consequently, if there are two conducting plates separated by a distance $y$,
then the only wave vectors in the $y$-direction that contribute to ${\bf E}$ and ${\bf B}$ 
are those which satisfy $k_y=n\pi/y$ (for $n=1,2,3,\ldots)$. In this way ${\bf E}$ and ${\bf B}$ attain 
a dependence on $y$, which we write as ${\bf E}\rightarrow \tilde{{\bf E}}(y)$, ${\bf B}\rightarrow \tilde{{\bf B}}(y)$.
Consequently we have $H_{\rm em} \rightarrow \tilde{H}_{\rm em}(y)$, which leads to a $y$-dependent vacuum energy
\begin{equation}\label{H2}
\tilde{E}_{\rm vac}(y)=\langle 0|\tilde{H}_{\rm em}(y)| 0\rangle .
\end{equation}
(As shown in Appendix \ref{APP1}, the vacuum $| 0\rangle$ can be considered as a state which does 
not depend on $y$.)
The general principles of classical mechanics then suggest that there should be the force between the plates
given by 
\begin{equation}\label{H3}
\tilde{F}(y)=-\frac{\partial \tilde{E}_{\rm vac}(y)}{\partial y} .
\end{equation}
A detailed calculation \cite{itz-zub} shows that $\tilde{E}_{\rm vac}(y)=\tilde{E}_{\rm fin}(y)+E_0$,
where $\tilde{E}_{\rm fin}(y)$ is finite and $E_0$ is an infinite constant which does not depend on $y$.
In this way (\ref{H3}) gives a finite result that turns out to agree with measurements \cite{lamoreaux}. 

The central idea of this paper is to question the validity of Eq.~(\ref{H3}), the equation 
which in the existing 
literature is usually taken for granted without further scrutiny. If (\ref{H3}) is valid, then
$y$ must be a dynamical variable with a kinetic part in the full Hamiltonian. Treating plates as 
classical non-relativistic objects, the minimal Hamiltonian that leads to (\ref{H3}) is
\begin{equation}\label{H4}
 \tilde{H}=\frac{p_y^2}{2m}+\tilde{E}_{\rm vac}(y) ,
\end{equation}
where $m=m_1m_2/(m_1+m_2)$ is the reduced mass of the two plates with masses $m_1$ and $m_2$.

From the point of view of general principles of classical mechanics \cite{LL1,goldstein}, however,
there is something suspicious about (\ref{H3}) and (\ref{H4}). The term $\tilde{E}_{\rm vac}(y)$ 
in (\ref{H4}) creates 
the force (\ref{H3}) on the dynamical variable $y(t)$ owing to the fact that 
$\tilde{E}_{\rm vac}(y)$ depends on $y$. On the other hand, $\tilde{E}_{\rm vac}(y)$ originates 
from (\ref{H1}) which does {\em not} depend on $y$, so $H_{\rm em}$ cannot generate any force 
on $y$ because $\partial H_{\rm em}/\partial y=0$. This looks like a paradox; how can it be that
$H_{\rm em}$ both depends and does not depend on $y$?

The answer, of course, is that $H_{\rm em}$ does not have any {\em explicit} dependence on 
$y$. Yet $\tilde{H}_{\rm em}(y)$ has an {\em implicit} dependence on $y$, i.e. the 
dependence which originates from using solutions of the equations of motion.

Now what kind of $y$-dependence is responsible for forces, explicit or implicit?
The general principles of classical mechanics \cite{LL1,goldstein} tell us that it is 
only explicit dependence that counts! Consequently, {\em (\ref{H3}) is not a legitimate 
formula to calculate the force}. 

To avoid misunderstanding, it is important to emphasize that we do not claim that 
(\ref{H3}) is wrong in a phenomenological sense. Phenomenologically, it may give 
a correct result. All we claim is that (\ref{H3}) is {\em not fundamental},
i.e. does not follow directly from {\em general principles}.

In this way we see that our critique of Eq.~(\ref{H3}) has the origin in 
general principles of classical mechanics. Nevertheless, in this paper we shall perform a full 
quantum analysis and we shall see that a similar critique of (\ref{H3}) works also in quantum physics.
The essence of the error committed in (\ref{H3}) is treating the implicit dependence 
as if it was explicit, which is equally illegitimate in both classical and quantum theory,
even if sometimes leads to phenomenologically correct results.

\section{The main proof}

The full action of quantum electrodynamics (QED) can be written as
\begin{equation}\label{S}
 I=I_{\rm em}(A)+I_{\rm matt}(\phi)+I_{\rm int}(A,\phi) ,
\end{equation}
where $A(x)=\{ A^{\mu}(x) \}$ is the EM field and $\phi(x)$ denotes all matter fields.
(The same notation $X(Y)$ is used for both functions such as $\phi(x)$ and functionals 
such as $I_{\rm em}(A)$. The meaning of this uniform notation in each particular case should be clear 
from the context.)
Explicitly
\begin{equation}\label{Sem}
I_{\rm em}(A)=-\frac{1}{4}\int d^4x\, F^{\mu\nu}F_{\mu\nu} , 
\end{equation}
\begin{equation}\label{Sint}
I_{\rm int}(A,\phi)=-\int d^4x\, A_{\mu}j^{\mu}(\phi) , 
\end{equation}
where $F_{\mu\nu}=\partial_{\mu}A_{\nu}-\partial_{\nu}A_{\mu}$ and $j^{\mu}(\phi)$ is the charge current.
The explicit expressions for $I_{\rm matt}(\phi)$ and $j^{\mu}(\phi)$ will not be needed.

The action (\ref{S}) can be written in terms of a Lagrangian $L$ as 
\begin{equation}
 I(A,\phi)=\int dt\, L(A,\dot{A},\phi,\dot{\phi}) ,
\end{equation}
where the dot denotes the time derivative.
Then by standard canonical methods \cite{weinberg} one can transform Lagrangian $L$ into the 
Hamiltonian 
\begin{equation}
 H=H_{\rm em}(A,\pi_A)+H_{\rm matt}(\phi,\pi_{\phi})+H_{\rm int}(A,\phi,\pi_{\phi}) ,
\end{equation}
where $\pi_A$ and $\pi_{\phi}$ are the canonical momenta and
each of the 3 terms is generated by the corresponding term in (\ref{S}).  

The time evolution of quantum variables is governed by the Hamiltonian. We use the Heisenberg picture,
so matter variables $\phi$ and $\pi_{\phi}$ obey the Heisenberg equations of motion
\begin{equation}\label{heis}
 \dot{\phi}=i[H,\phi], \;\;\; \dot{\pi}_{\phi}=i[H,\pi_{\phi}] .
\end{equation}
In particular, the quantity $i[H,\pi_{\phi}]$ on the right-hand side of the second equation
in (\ref{heis}) describes the quantum force on the matter. 
We stress that (\ref{heis}) is {\em exact}, so {\em all quantum forces} on matter that can be derived from QED 
are described by (\ref{heis}). 
In particular, (\ref{heis}) is a non-perturbative result, so in principle it
contains effects of all higher order quantum loop diagrams. Of course, the contributions of loop diagrams
are not explicit because we work in the non-perturbative Heisenberg picture. Loop diagrams 
are a perturbative concept, so they could be seen explicitly if we worked in the Dirac interaction picture. 
   
Now the crucial observation is the fact that $H_{\rm em}(A,\pi_A)$ does not have any explicit dependence on
$\phi$ and $\pi_{\phi}$. Consequently
\begin{equation}\label{comm1}
 [H_{\rm em},\phi]=0, \;\;\; [H_{\rm em},\pi_{\phi}]=0 ,
\end{equation}
so (\ref{heis}) reduces to
\begin{eqnarray}\label{heis_reduced}
& \dot{\phi}=i[H_{\rm int},\phi]+i[H_{\rm matt},\phi] , &
\nonumber \\ 
& \dot{\pi}_{\phi}=i[H_{\rm int},\pi_{\phi}] + i[H_{\rm matt},\pi_{\phi}] . &
\end{eqnarray}
Thus we see that all quantum forces on matter are generated by $H_{\rm int}$ and $H_{\rm matt}$.
In other words, we have proven that {\em $H_{\rm em}$ does not have any contribution to the quantum force on matter.}

Similarly, instead of the canonical momentum $\pi_{\phi}$, one can study the kinematic momentum of matter
\begin{equation}
 P^i(\phi)=\int d^3x\, T_0^i(\phi) ,
\end{equation}
where $T_{\mu}^{\nu}(\phi)$ is the energy-momentum tensor of matter. Since $P^i(\phi)$ does not have 
an explicit dependence on EM fields, it follows that  
\begin{equation}
 [H_{\rm em},P^i]=0. 
\end{equation}
Consequently
\begin{equation}
\dot{P^i} = i[H_{\rm int},P^i] + i[H_{\rm matt},P^i] ,
\end{equation}
which shows that $H_{\rm em}$ does not contribute to the forces on matter in the sense of 
changing the kinematic momentum of matter. 

So far our analysis was very general and we said nothing specific about the vacuum energy of EM field. 
To see the role of vacuum energy consider a state of the form 
\begin{equation}\label{Omega}
|\Omega\rangle=|0_A\rangle|\psi_{\phi}\rangle, 
\end{equation}
where $|0_A\rangle$ is the EM vacuum and $|\psi_{\phi}\rangle$
%
is some matter state. In general $|\psi_{\phi}\rangle$ can be an arbitrary physical state,
but in a study of Casimir force one takes $|\psi_{\phi}\rangle$ which corresponds
to physically realistic conducting plates, with well defined boundaries on which 
EM fields satisfy appropriate boundary conditions. Similarly to the vacuum discussed
in Appendix \ref{APP1}, the state $|\Omega\rangle$ can be represented either as a state 
in a larger Hilbert space of all possible physical states (most of which
have nothing to do with Casimir plates), or, equivalently, as a state in a smaller Hilbert subspace
containing only those states which can describe given Casimir plates.  
In the latter representation the matter state can be viewed as a state with a parametric dependence 
on EM fields (similar to the $y$-dependence in Appendix \ref{APP1}), but in the former representation
such a dependence is absent. In the rest of the discussion we adopt the representation 
without the parametric dependence, which seems to be more elegant, more systematic, and more general.

In the EM vacuum we have $\langle 0_A|A^{\mu}|0_A \rangle=0$, 
so using the fact that $H_{\rm int}$ is linear in $A^{\mu}$ due to (\ref{Sint}), we have
\begin{equation}
 \langle\Omega |H_{\rm int}| \Omega\rangle =0.
\end{equation}
On the other hand $H_{\rm em}$ is quadratic in $A^{\mu}$ due to (\ref{Sem}), so
all EM vacuum energy comes from the term 
\begin{equation}\label{Evac}
 E_{\rm vac}=\langle\Omega |H_{\rm em}| \Omega\rangle=\langle 0_A |H_{\rm em}| 0_A\rangle.
\end{equation}
Of course, the term $\langle\Omega |H_{\rm matt}| \Omega\rangle=\langle \psi_{\phi} |H_{\rm matt}| \psi_{\phi}\rangle$ 
also contributes to the total energy in the EM vacuum, 
but this is clearly the matter energy determined by the matter state $| \psi_{\phi}\rangle$,
so it cannot be considered as a contribution to the EM vacuum energy.

To conclude, from (\ref{heis_reduced}) we see that $H_{\rm em}$ does not contribute to quantum forces on matter,
and from (\ref{Evac}) we see that only $H_{\rm em}$ contributes to the EM vacuum energy. 
This proves our main result that {\em EM vacuum energy does not contribute to quantum forces on matter}.

In principle our paper could stop here. Nevertheless it may be illuminating 
to consider some additional issues. If, as we just proved, 
vacuum energy does not create forces on matter, then what is wrong 
with Eq. (\ref{H3})? In the rest of the paper we 
provide a deeper understanding of the error committed in (\ref{H3}).

\section{An illegitimate use of the equations of motion}

In the Lorenz gauge $\partial_{\mu}A^{\mu}=0$ 
the equations of motion for EM field are
\begin{equation}\label{boxA}
 \Box A^{\mu}=j^{\mu}(\phi) ,
\end{equation}
with $\Box\equiv \partial^{\nu}\partial_{\nu}$. This equation of motion can be solved as
\begin{equation}\label{tildeA}
 A^{\mu}=\tilde{A}^{\mu}(\phi) ,
\end{equation}
where 
\begin{equation}
\tilde{A}^{\mu}(\phi)\equiv \Box^{-1}j^{\mu}(\phi).
\end{equation}
In this way the use of equations of motion allows one to express the EM field as a functional of matter fields.
Thus, even though $A^{\mu}$ does not have an {\em explicit} dependence on $\phi$,
it depends on $\phi$ {\em implicitly} due to (\ref{tildeA}). 
Inserting (\ref{tildeA}) into the expression for $H_{\rm em}(A,\pi_A)$ one gets the quantity
\begin{equation}
 \tilde{H}_{\rm em}(\phi)=H_{\rm em}(\tilde{A}(\phi),\tilde{\pi}_A(\phi)) .
\end{equation}
In this way one gets non-vanishing commutators
\begin{equation}\label{comm2}
 [\tilde{H}_{\rm em},\phi]\neq 0, \;\;\; [\tilde{H}_{\rm em},\pi_{\phi}]\neq 0 ,
\end{equation}
despite the fact that the commutators (\ref{comm1}) vanish. 

So far we did nothing wrong. But now consider the following step. Guided by the correct 
Heisenberg equations of motion (\ref{heis}), one may be tempted to write
\begin{equation}\label{heis2}
 \dot{\pi}_{\phi}\stackrel{?}{=}i[\tilde{H}_{\rm em},\pi_{\phi}] .
\end{equation}
For the moment let us pretend that (\ref{heis2}) was legitimate, to see where that would lead us. 
If that was legitimate, then the quantity $i[\tilde{H}_{\rm em},\pi_{\phi}]$ would represent a 
non-vanishing force generated by $\tilde{H}_{\rm em}$. In the EM vacuum state (\ref{Omega}) we would have
\begin{equation}\label{heis3}
 \langle\Omega|\dot{\pi}_{\phi}|\Omega\rangle =i \langle\Omega|[\tilde{H}_{\rm em},\pi_{\phi}]|\Omega\rangle .
\end{equation}
so we would get a force that originates from the EM vacuum. When $\phi$ is a bosonic field,
the momentum operator can be represented as
$\pi_{\phi}=-i\delta/\delta\phi$, so (\ref{heis2}) can be written as
\begin{equation}\label{heis4}
 \dot{\pi}_{\phi}=-\frac{\delta\tilde{H}_{\rm em}}{\delta\phi} .
\end{equation}
To get the Casimir force one can model matter field $\phi({\bf x},t)$ with a single degree of freedom $y(t)$ 
representing the distance between the Casimir plates, in which case (\ref{heis4}) reduces to
\begin{equation}\label{heis5}
 \dot{p}_y=-\frac{\partial\tilde{H}_{\rm em}}{\partial y} .
\end{equation}
Hence in the EM vacuum we have
\begin{equation}\label{heis6}
 \langle\Omega|\dot{p}_y|\Omega\rangle=-\langle\Omega|\frac{\partial\tilde{H}_{\rm em}}{\partial y}|\Omega\rangle .
\end{equation}
The distance between the plates $y(t)$ is a macroscopic observable, so it can be approximated by a classical variable
$y_c(t)=\langle\Omega|y(t)|\Omega\rangle$. As a consequence, (\ref{heis6}) can be approximated by
\begin{equation}\label{heis7}
 \dot{p}_{y,c}=-\frac{\partial\langle\Omega|\tilde{H}_{\rm em}|\Omega\rangle}{\partial y_c} .
\end{equation}
The approximation which gives (\ref{heis7}) from (\ref{heis6}) is the same approximation that is used in elementary
quantum mechanics to derive the classical equations of motion from the Ehrenfest theorem \cite{cohen-tann1,basd-dalib}.
(Of course, it does not mean that Casimir force can be explained by classical physics, because 
here only the matter degrees described by $y_c$ are described classically, while EM degrees are still 
treated quantum mechanically.) 
Eq.~(\ref{heis7}) is equivalent to Eq.~(\ref{H3}), here obtained from a more general equation of motion (\ref{heis2}).
In the derivation of (\ref{H3}) the variable $y$ was treated as a classical variable from the beginning,
which, of course, is an approximation too. In this sense $y$ in (\ref{H3}) corresponds to 
the classical variable $y_c$ in (\ref{heis7}), not to the quantum variable $y$ in (\ref{heis6}).

Finally, from a formal point of view it is interesting to see what happens if all microscopic variables 
are considered in the classical limit. In that limit all the commutators turn into Poisson brackets according to the rule
$[A,B]\rightarrow i\{A_c,B_c\}$, so (\ref{heis2}) turns into
\begin{equation}\label{heis8}
 \dot{\pi}_{\phi,c}=\{ \pi_{\phi,c},\tilde{H}_{{\rm em},c} \} 
=-\frac{\delta\tilde{H}_{{\rm em},c}}{\delta\phi_c} .
\end{equation}
This classical equation has the same form as the quantum equation (\ref{heis4}).

Nevertheless, (\ref{heis2}) was not legitimate. Consequently, all equations (\ref{heis3})-(\ref{heis8}) above 
were illegitimate. When calculating commutators in canonical equations of motion 
in the Heisenberg picture, one must use $H$, not $\tilde{H}$. 
In other words, to calculate the commutators in the Heisenberg equations of motion, what counts is the 
explicit dependence \cite{dirac,schiff,cohen-tann1,merzbacher,basd-dalib}, 
not the implicit dependence obtained by solving the equations of motion.
In general, the use of equations of motion in such a way leads to wrong results.
Next we demonstrate this fact explicitly, by solving a simple toy model.

\section{A toy model}

To see what goes wrong when equations of motion are used 
in a way described above, let us study a concrete example.
Instead of field $A^{\mu}({\bf x},t)$ with an infinite number of degrees of freedom,
let us consider a single degree of freedom $A(t)$. Similarly, instead of matter field $\phi({\bf x},t)$,
let us consider another single degree of freedom $y(t)$. Let the dynamics of these two degrees of freedom
be described by the Hamiltonian 
\begin{equation}
 H=H_A+H_y+H_{\rm int} ,
\end{equation}
where
\begin{equation}\label{toy2}
 H_A=\frac{p_A^2}{2} , \;\;\; H_y=\frac{p_y^2}{2}, \;\;\;  H_{\rm int}=-\gamma^2 Ay ,
\end{equation}
and $\gamma^2$ is a coupling constant. A straightforward calculation shows that the resulting equations of motion are
\begin{equation}\label{toy3}
 \ddot{A}=F_A , \;\;\; \ddot{y}=F_y ,
\end{equation}
where the forces are given by
\begin{equation}\label{toy4}
 F_A=-\frac{\partial H}{\partial A}=\gamma^2 y, \;\;\;  F_y=-\frac{\partial H}{\partial y}=\gamma^2 A .
\end{equation}

Let us study one particular solution of (\ref{toy3})
\begin{equation}\label{toy5}
 A(t)=e^{-\gamma t} , \;\;\; y(t)=e^{-\gamma t} .
\end{equation}
Inserting the solution into the second equation in (\ref{toy4}) we obtain
\begin{equation}\label{toy6}
 F_y=\gamma^2 y ,
\end{equation}
which is the correct result.

Now using $p_A=\dot{A}$, the first equation in (\ref{toy2}) can be written as 
\begin{equation}\label{toy7}
H_A=\frac{\dot{A}^2}{2} .
\end{equation}
Using the solution (\ref{toy5}), we can write $\dot{A}=\dot{y}=-\gamma y$. 
Inserting this into (\ref{toy7}), we get
\begin{equation}\label{toy8}
 \tilde{H}_A(y)=\frac{\gamma^2 y^2}{2} .
\end{equation}
From this one may be tempted to calculate the force as
\begin{equation}\label{toy9}
 \tilde{F}_y=-\frac{\partial \tilde{H}_A}{\partial y}=-\gamma^2 y .
\end{equation}
However, comparing (\ref{toy9}) with (\ref{toy6}) one can see very explicitly that $\tilde{F_y}\neq F_y$. 
This explicitly demonstrates that {\em (\ref{toy9}) is not a legitimate way to calculate the force}. 

In the simple case above the forces (\ref{toy9}) and (\ref{toy6}) turn out to have the same absolute value 
and the opposite sign (the reader is encouraged to check all the signs by himself/herself),
but this is not a general rule. The only rule is that $\tilde{F_y}$ and $F_y$ are, in general, different.
It is not excluded that, in some cases, $\tilde{F_y}$ and $F_y$ turn out to give the same result,
but this is an exception rather than a rule.

\section{Discussion and conclusion}

We have shown 
that calculating Casimir force by (\ref{H3}) is not legitimate. 
Yet, it is known that such a calculation leads to a result which is in agreement with experiments, as well
as with results obtained by other methods that do not involve vacuum energy. 
The methods with and without vacuum energy give the same results not only for perfect conductors,
but even for realistic materials (compare e.g. \cite{zhou-spruch} with \cite{tomas}).
How can it be that an illegitimate method leads to a correct result?

Our analysis based on Heisenberg picture, which is very suitable for proving general results,
cannot easily answer such a more specific question. This question can be answered by a very different
method \cite{jaffe}, based on perturbative expansion in Feynman diagrams. The force calculated 
in terms of vacuum energy turns out not to be an exact result, but an approximation 
corresponding to the limit in which
fine structure constant goes to infinity \cite{jaffe}. In this sense Casimir force, like 
any other force in quantum field theory, originates from the interaction term $H_{\rm int}$.    
More precisely, as stressed in \cite{jaffe}, the physical origin of Casimir force 
lies in the van der Waals forces between the conducting plates.

Similarly to the vacuum energy, the van der Waals forces also originate from quantum fluctuations.
However, the important difference lies in the fact that van der Waals forces do not originate
from {\em vacuum} fluctuations. The van der Waals force originates from matter fluctuations 
of charge density \cite{cohen-tann2,parsegian}, but it does not depend on the operator ordering
of charge current operator $j^{\mu}(\phi)$ in the interaction term $L_{\rm int}=-\int d^3x\, A_{\mu}j^{\mu}(\phi)$.
The force persists even if $j^{\mu}(\phi)$ is normal ordered so that 
$\langle 0_{\phi}|j^{\mu}(\phi)|0_{\phi}\rangle =0$.

As a side remark let us also note that the charge current {\em must} be normal ordered for physical reasons. 
Without normal
ordering one would have $\langle 0_{\phi}|j^{\mu}(\phi)|0_{\phi}\rangle \neq 0$ \cite{nik-grg}, so the Maxwell 
equation
\begin{equation}
 \partial_{\mu}F^{\mu\nu}=j^{\nu}
\end{equation}
would imply existence of EM fields even in the vacuum, in a clear contradiction with observations. 
We also note that taking normal ordering of $j^{\mu}$ in the Maxwell equation is analogous 
to taking normal ordering in the energy-momentum tensor $T^{\mu\nu}$ in the Einstein equation, 
the latter being closely related to the cosmological constant problem 
in gravitational physics \cite{weinberg_cc,carroll_cc,padm_cc,nobb_cc}. 

To conclude, in this paper we have shown that the standard calculation of Casimir force 
from the vacuum energy through Eq.~(\ref{H3}) is illegitimate. Essentially, this is because the vacuum energy 
has only implicit dependence on the distance $y$ between the plates, namely the dependence that
originates from the solution of the equations of motion, while the legitimate application of (\ref{H3}) 
would require an explicit dependence on $y$, which is absent. Therefore, at the fundamental level,
the Casimir force does not originate from the vacuum energy. This is in full accordance 
with the result by Jaffe \cite{jaffe}, according to which the true physical origin 
of the Casimir force lies in the non-vacuum van der Waals forces between the material plates.   

\section*{Acknowledgements}

The author is grateful to M. Cerdonio,
T. Padmanabhan and M.S. Toma\v{s} for discussions,
and to an anonymous referee for objections that stimulated further clarifications. 
This work was supported by the Ministry of Science of the Republic of Croatia
and partially supported by the H2020 Twinning project No. 692194, ``RBI-T-WINNING''.

\appendix

\section{Parametric (in)dependence of the vacuum state}
\label{APP1}

\subsection{Free field and free vacuum}

Consider the free EM field, i.e. the situation in which Casimir plates are {\em not} present. 
In this case the EM field can be expanded as
\begin{equation}\label{ap2}
A^{\mu}(x)=\sum_P a_P \epsilon_P^{\mu} e^{-ipx} +{\rm h.c.} ,
\end{equation}
where $x=(x^0,x^1,x^2,x^3)$ is a spacetime point,
$a_P$ are destruction operators and $\epsilon_P^{\mu}$ are polarization vectors.
Here we use a compact notation
\begin{equation}
 P=(p,\lambda) ,
\end{equation}
where $\lambda=1,2$ are polarizations of the EM field and $p=(|{\bf p}|, {\bf p})=(|{\bf p}|, p^1,p^2,p^3)$ 
are 4-momenta. When the system lives in an infinite volume, then $p^1,p^2,p^3$ are continuous variables, in which case 
the symbol $\sum_P$ means 
\begin{equation}\label{aps1}
 \sum_P = \int \frac{d^3 p}{(2\pi)^3} \sum_{\lambda} .
\end{equation}
Alternatively, when the system lives in a very large but finite volume $V$, then $p^1,p^2,p^3$ are quasi-continuous
(i.e. discrete with a very small spacing in the momentum space), in which case 
the symbol $\sum_P$ means
\begin{equation}\label{aps2}
 \sum_P = \frac{1}{V}\sum_{{\bf p}} \sum_{\lambda} .
\end{equation}
To emphasize physics rather than mathematics, in the rest of the discussion we shall not attempt 
to be fully mathematically rigorous. In particular we shall not state explicitly
whether (\ref{aps1}) or (\ref{aps2}) is used. Let us only note that some formal manipulations
that we shall perform can be much more easily defined rigorously when (\ref{aps2}) is understood.

Now let us define the vacuum $|0\rangle$ in a standard way, by requiring that 
\begin{equation}\label{ap4}
 a_P|0\rangle =0 
\end{equation}
for all $P$.
Next let us make an artificial split
\begin{equation}\label{ap_split}
\sum_P = \sum_K + \sum_Q
\end{equation}
where $K=(k,\lambda)$ and $Q=(q,\lambda)$ are defined by
\begin{equation}\label{ap_kq}
k^2=n\pi/y , \;\;\; q^2\neq n\pi/y .
\end{equation}
Here $k^2$ and $q^2$ are components of the 4-momenta $k$ and $q$ in the $x^2$-direction, 
while $y$ is a single parameter of the dimension of length. 
Eq.~(\ref{ap_kq}) means that $k^2$ takes only discrete values with $n=1,2,3,\ldots$,
while $q^2$ takes all the other values except those discrete ones.
With the artificial split (\ref{ap_split}), the free field (\ref{ap2}) can be rewritten as
\begin{equation}\label{ap2r}
A^{\mu}(x)=\sum_K a_K \epsilon_K^{\mu} e^{-ikx} + \sum_Q a_Q \epsilon_Q^{\mu} e^{-iqx}  +{\rm h.c.} ,
\end{equation}
while (\ref{ap4}) can be written as 
\begin{equation}\label{ap4r}
 a_K|0\rangle =0, \;\;\;  a_Q|0\rangle =0 .
\end{equation}
Note that the two terms in (\ref{ap2r}), one involving $\sum_K$ and another involving $\sum_Q$,
both depend on $y$. Nevertheless, their sum involving  $\sum_K + \sum_Q$ does not depend on $y$.
Similarly, each of the two equations in (\ref{ap4r}) depends on $y$, but together
they are equivalent to (\ref{ap4}) which does not depend on $y$.

\subsection{Field and vacuum in the presence of Casimir plates}

The artificial split above is useful for the sake of comparison with fields in the presence of Casimir plates.
When Casimir plates separated by a distance $y$ in the $x^2$-direction 
are present, instead of (\ref{ap2r}) we have
\begin{equation}\label{ap2c}
A^{\mu}(x;y)=\sum_K a_K \epsilon_K^{\mu} e^{-ikx}  +{\rm h.c.} .
\end{equation}
We see that the field (\ref{ap2c}) has an extra parametric dependence on $y$, not shared by  (\ref{ap2r}).
Does the corresponding vacuum $|0'\rangle$ also attains an extra parametric dependence on $y$? At first sight the answer 
seems to be ``yes'', because now it seems natural to take $|0'\rangle=|0(y)\rangle$, where $|0(y)\rangle$
is defined by
\begin{equation}\label{ap4c}
 a_K|0(y)\rangle =0 .
\end{equation}
However, with such a definition, one might ask what is $a_Q|0'\rangle$?
There are two possible answers, corresponding to two possible 
representations of the field algebra. 

One possible answer is that $|0(y)\rangle$ lives in a Hilbert space
which is only a subspace of the Hilbert space in which $|0\rangle$ lives. 
The operator $a_Q$ is not defined on this smaller Hilbert space, so the quantity $a_Q|0(y)\rangle$
is simply not defined. 

Another possible answer is that the vacuum $|0'\rangle$ should live in the same Hilbert 
space as $|0\rangle$. Therefore, in addition to the obvious requirement $a_K|0'\rangle=0$, 
the quantity $a_Q|0'\rangle$ must also be well defined. For definiteness one can take
$a_Q|0'\rangle=0$, so instead of (\ref{ap4c}) we have
\begin{equation}\label{ap4rn}
 a_K|0'\rangle =0, \;\;\;  a_Q|0'\rangle =0 .
\end{equation}
Comparing it with (\ref{ap4r}) we see that $|0'\rangle$ is essentially the same as $|0\rangle$,
differing from it at most by a physically irrelevant phase. Ignoring the phase, we can write $|0'\rangle=|0\rangle$.

Mathematically, the existence of two representations (one in a smaller Hilbert space than the other) of the vacuum 
is not much different from the fact that the vector 
$(1,0)$ in a 2-dimensional space can also be represented as $(1,0,0,0,\ldots)$ in a higher dimensional space. 
But which of the two representations is physically correct? 
They both are, because they are {\em physically equivalent}. Namely, physical observables (such as Hamiltonian)
in the presence of Casimir plates are built from the field operator
(\ref{ap2c}), which does not contain the operators $a_Q$ and $a^{\dagger}_Q$. Therefore, in calculation
of physical quantities, those operators never act on $|0'\rangle$, so it is physically irrelevant
what $a_Q|0'\rangle$ is. Since both definitions of the vacuum satisfy $a_K|0'\rangle =0$, it means
that the two definitions agree in the physically relevant sector, so they are physically equivalent. 

The equivalence of the two representations means that matrix elements such as 
$\langle 0'|\tilde{H}_{\rm em}(y)| 0'\rangle$ do not depend on representation.
In the representation (\ref{ap4rn}), the vacuum $| 0'\rangle=| 0\rangle$ does not depend on $y$, so in this 
representation it is manifest that only $\partial\tilde{H}_{\rm em}(y)/\partial y$ 
contributes to the right-hand side of (\ref{H3}), while $\partial|0'\rangle/\partial y=0$. 

In the representation (\ref{ap4c}) we have $\partial|0(y)\rangle/\partial y \neq 0$. However, we repeat
that this representation lives in a smaller Hilbert space. In particular, even though the vector
$|0(y)\rangle$ belongs to this small space, the derivative
\begin{equation}
\frac{\partial|0(y)\rangle}{\partial y} = \frac{|0(y+dy/2)\rangle - |0(y-dy/2)\rangle}{dy}
\end{equation}
does {\em not} belong to this small space. This is because $|0(y+dy/2)\rangle$ lives in {\em another}
small Hilbert space (the Hilbert space in which the distance between the plates is $y+dy/2$ rather than 
$y$), which is different from the small Hilbert space in which $|0(y)\rangle$ and $\tilde{H}_{\rm em}(y)$
live. Therefore, even though $\partial|0(y)\rangle/\partial y \neq 0$, 
the quantity $\tilde{H}_{\rm em}(y) \partial|0(y)\rangle/\partial y $ is not well defined,
so it cannot contribute to the right-hand side of (\ref{H3}). 

To conclude, we see that the vacuum in the presence of Casimir plates can be considered as a state which does 
not have a parametric dependence on $y$. In one representation this is because the vacuum is explicitly 
$y$-independent, while in another representation the mathematical dependence of the vacuum on $y$
does not have any physical consequences. Loosely speaking, the dependence of the vacuum on $y$ only 
determines the size of the smaller Hilbert space on which the vacuum is represented, while the physical content
of the vacuum state does not depend on it. The mathematics looks different in the two representations,
but physics is the same. Note, however, that the field operators (\ref{ap2r}) and (\ref{ap2c})
are really different in a physical sense, and their physical difference does not depend on the representation.

\end{document}